\documentclass[reprint,aps,prl,superscriptaddress,twocolumn,longbibliography,floatfix,showkeys,showpacs]{revtex4-2}

\usepackage{stix2}

\usepackage[english]{babel}
\usepackage[autostyle=true]{csquotes}

\usepackage{mathtools}
\usepackage{amssymb}
\usepackage{booktabs}
\newcommand{\ra}[1]{\renewcommand{\arraystretch}{#1}}
\usepackage{graphicx}
\usepackage{siunitx}
\usepackage{ulem}
\usepackage{bbold}

\usepackage[unicode=true,
 bookmarks=true,bookmarksnumbered=true,bookmarksopen=true,bookmarksopenlevel=2,
 breaklinks=false,pdfborder={0 0 1},backref=false,colorlinks=true]
 {hyperref}
\hypersetup{linkcolor=blue, citecolor=blue, urlcolor=blue, filecolor=blue, pdfpagelayout=OneColumn, pdfnewwindow=true, pdfstartview=XYZ, plainpages=false}
\usepackage[all]{hypcap}

\begin{document}

\title{Equivalence of light transport and depolarization}

\author{Maximilian Gill}
\affiliation{1. Physikalisches Institut, Universit\"at Stuttgart, Pfaffenwaldring 57, 70569 Stuttgart, Germany}
\author{Bruno Gompf}
\email{b.gompf@pi.uni-stuttgart.de}
\affiliation{1. Physikalisches Institut, Universit\"at Stuttgart, Pfaffenwaldring 57, 70569 Stuttgart, Germany}
\author{Martin Dressel}
\affiliation{1. Physikalisches Institut, Universit\"at Stuttgart, Pfaffenwaldring 57, 70569 Stuttgart, Germany}
\author{Gabriel Schnoering}
\email{schngabr@ethz.ch}
\affiliation{1. Physikalisches Institut, Universit\"at Stuttgart, Pfaffenwaldring 57, 70569 Stuttgart, Germany}
\affiliation{Laboratory of Thermodynamics in Emerging Technologies, ETH Zurich, Sonneggstrasse 3, Zurich, Switzerland}



\maketitle

\begin{bf}
The study of scattered polarized light has led to important advances in distinct fields such as astronomy, atmospheric sciences and bio-imaging \cite{Hough_2006, Akiyama_2021first, Sorensen_2001, Roggemann_2018, Ntziachristos_2010, Ghosh_2011}. In random diffusing media, light disorientation and the scrambling of its polarization state appear to always occur together \cite{Bicout_1994, Brosseau_1994}. Their apparent inseparability suggests a profound connection between optical transport and depolarization. Here, we present experimental evidence of their equivalence and quantify their relationship in colloidal suspensions of microscopic constituents. In particular, a proportionality relation between optical transport lengths and their depolarization counterparts is provided. This equivalence imposes depolarization whenever light traverses random media and holds for wide spectral ranges and scatterer concentrations. Our results clarify the connection between microscopic processes and measurable polarization signatures.
\end{bf}

\begin{figure}[b]
  \centering
  \includegraphics[width=89mm]{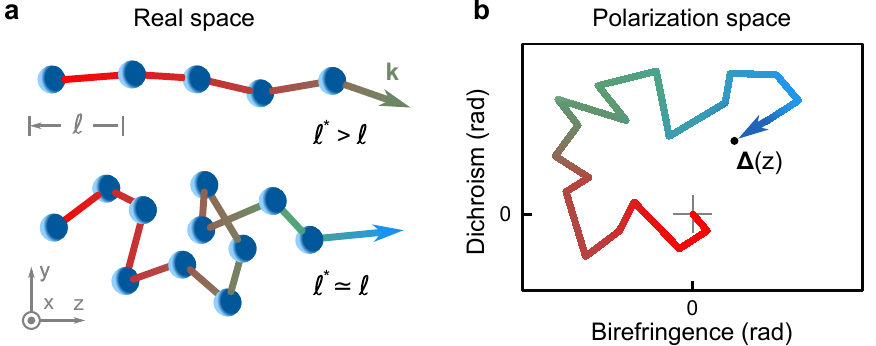}
  \caption {\textbf{a}, Sketch of the randomization of $\mathbf{k}$ in real space for two different ratios of $\ell^* / \ell$. A shorter transport mean free path $\ell^*$ implies faster randomization.
    \textbf{b}, A random walk of the optical property $\mathbf{\Delta}$ in the birefringence/dichroism plane represents the randomization of polarization (see eq. $\eqref{eq:BK}$).}
  \label{fig:poldiff}
\end{figure}

In complex media, light diffusion is governed by two length scales, the mean free path $\ell$ describing the interspace between two scattering events, and the transport mean free path $\ell^*$ corresponding to the distance that light travels until its direction $\mathbf{k}$ is randomized, as illustrated in Fig.~$\ref{fig:poldiff}\mathrm{(a)}$ \cite{Rojas_Ochoa_2002}. Randomization of the propagation direction of light is common in cold atomic clouds \cite{Labeyrie_2003}, semiconductors \cite{Gomez_Rivas_2001}, interstellar dust \cite{Weingartner_2001}, clouds \cite{Hansen_1971}, biological tissues \cite{Profio_1989, Liu_2018} and turbid media \cite{Weitz_1993}. There, light diffusion can be singular \cite{Patsyk_2020}, controlled \cite{Rotter_2017, de_Aguiar_2017} or propagate with constant time \cite{Savo_2017}.

Complementing spectral and imaging methods, polarization measurements reveal valuable information about the medium traversed by light \cite{Fujiwara_2007, Azzam_1978, Ossikovski_2011, Agarwal_2015}. Accurate models that describe depolarization by multiple scattering are complex \cite{Pattelli_2018, Li_2018, Popoff_2010, Canaguier_Durand_2020} and the ability to quantify it in a simple way would provide great opportunities. Although depolarization is a macroscopic observable, generally acquired over long integration times, it gives access to microscopic parameters and processes. These usually occur on short time scales and are hard to measure by other means \cite{Rojas_Ochoa_2002}. Examples are refractive indices, geometries, distances between and motion of microscopic defects and dynamic light transport quantities $(\ell,~\ell^*)$ \cite{Bicout_1994, Brosseau_1994, Rojas_Ochoa_2002}. Furthermore, advances in the understanding of depolarization open exciting new pathways to explore the optics of complex particulate matter.

Stochastic modelling approaches successfully describe multiple scattering random media \cite{Bicout_1994, Devlaminck_2015, Charbois_2016}. Early observations showed a linear evolution of depolarization quantities at large penetration depths $z$ that agreed well with Monte Carlo simulations \cite{Bicout_1994}. Stochastic differential equations analytically connect polarization diffusion to measured depolarization values and provide a deeper picture about contributing phenomena \cite{Charbois_2016}. Depolarization propagation is associated with two regimes: a linear evolution at large $z$ and a non-linear transient regime at short $z$, each with its distinct polarization dynamics \cite{Charbois_2016}. Studies on Rayleigh and Mie scatterers have long suggested a relation between optical transport $(\ell,~\ell^*)$ and depolarization \cite{Bicout_1994}. Here, we show their equivalence and quantify their proportionality.

Using optical transport quantities in a random medium, the light trajectory can be modelled as an isotropic random walk with direction changes due to stochastic scattering processes. Its mean squared displacement $\sigma^2_r(t) = \langle|\mathbf{r}(t) - \mathbf{r}(0)|^2\rangle$ follows the equation \cite{Pierrat_2006}
\begin{equation}
\sigma^2_r(t) = Dt
\label{eq:photon}
\end{equation}
over time $t$, where $D$ is the diffusion coefficient defined by $D = c\ell^*/3$, $c$ is the phase-velocity of light in the medium and $\langle ... \rangle$ denotes averaging over the ensemble of possible light paths. 

Since the dynamics in real space are governed by characteristic length scales we expect the same to be true in polarization space. In analogy to the coordinate vector of light $\mathbf{r}$ in real space, a set of three 2-d vectors $\mathbf{\Delta}_b$, ($b \in \{\mathrm{L},\mathrm{L45},\mathrm{C} \}$) encode the accumulated polarization state of the light in the birefringence-dichrosim space (Fig.~$\ref{fig:poldiff}\mathrm{(b)}$). These elements exist in three different bases: linear (L), linear at $\pm 45^\circ$ (L45) and circular (C) \cite{Jones_1948} and change randomly at each scattering event. The mean squared polarization displacement,
\begin{equation}
\sigma^2_b(z) = \langle\vert \mathbf{\Delta}_b(z) - \mathbf{\Delta}_b(0) \vert^2\rangle,
\label{eq:sigmaDelta}
\end{equation}
defines the depolarization and depends on the light penetration depth $z$ into the medium \cite{Devlaminck_2015, Charbois_2016}. In analogy to eq. \eqref{eq:photon}, experiments have shown that it satisfies \cite{Bicout_1994, Charbois_2016}
\begin{equation}
  \lim_{z\gg 0}\sigma^2_b(z) = B_bz + K_b,
  \label{eq:BK}
\end{equation}
for large $z$. Comparison of eqs. \eqref{eq:photon} and \eqref{eq:BK} reveals that $B_{b}$ defines a diffusion coefficient for polarization states and its inverse $B^{-1}_{b}$ is the associated depolarization length scale.  The additional variance $K_b$ accounts for depolarization contributions earlier along the $z$-propagation. It is a testimony to the fact that the polarization diffusion $\sigma_b^2(z)$ exhibits different statistics at small penetration depths $0 \leq z \lesssim z^{\mathrm{T}}_b$ when compared to $\sigma^2_r(t)$. These contributions are short-lived and decay exponentially over a characteristic length scale $z_b^\mathrm{T}$, the so-called \textit{transient regime} \cite{Charbois_2016}. Previous studies found that $\sigma_b^2(z)$ evolves non-linearly in this regime \cite{Charbois_2016, Agarwal_2015}. This behaviour quickly recedes as $z$ enters the \textit{stationary regime} for $z \geq z^{\mathrm{T}}_b$. The two length scales $z_b^\mathrm{T}$ and $B_b^{-1}$ are prominent features of depolarization and allow analysing the different evolutions of $\sigma_b^2(z)$.

In order to determine $(z_b^\mathrm{T}, B_b^{-1})$, the entire depolarization curve $\sigma_b^2(z)$ is required. The random perturbation of $\mathbf{\Delta}_b(z)$ along the propagation direction ${z}$ can be described by an Ornstein-Uhlenbeck process \cite{Uhlenbeck_1930}, i.e. by a stochastic differential equation, whose coefficients encode microscopic dynamics. The accumulated optical property relates to its \textit{polarization velocity} $\mathbf{p}_b(z)$ through
\begin{equation}
  \mathbf{\Delta}_b(z) = \int_{0}^z\mathbf{p}_b(s)ds.
  \label{eq:Delta}
\end{equation}
$\mathbf{p}_{b}(z)$ satisfies the stochastic differential equation \cite{Oksendal_1998}:
\begin{align}
  \frac{\mathrm{d}\mathbf{p}_b(z)}{\mathrm{d}z} = \mathbf{A}_b\mathbf{p}_b(z) + \mathbf{\Sigma}_b\frac{\mathrm{d}\mathbf{N}_b(z)}{\mathrm{d}z},
  \label{eq:stoch}
\end{align}
with $\mathbf{N}_b(z)$ a unit normal noise vector, $\mathbf{\Sigma}_b$ a real $2 \times 2$ noise amplitude and $\mathbf{A}_{b}$ a real $2 \times 2$ drift matrix.

The dynamics of $\mathbf{p}_b(z)$ are not completely random, but originate from an interplay between random noise provided by $\mathbf{N}_b(z)$ and deterministic interchange and damping given by $\mathbf{A}_b$ (eq. (\ref{eq:stoch})). At short distances $z$, the term $\mathbf{A}_b\mathbf{p}_b(z)$ dominates. At large distances both terms drive the dynamics. For this reason depolarization $\sigma^2_b(z)$ experiences two distinct regimes at different scales of $z$ (eq. \eqref{eq:BK}).

The analytic expression for the $z$-dependent curves of $\sigma_b^2(z)$, derived from eqns. (\ref{eq:sigmaDelta},~\ref{eq:Delta},~\ref{eq:stoch}) (Supporting Information S2), is later fitted to the experimental data. The fits provide estimates for the coefficients $\mathbf{A}_b$ and $\mathbf{\Sigma}_b$, that define explicit expressions for $B_b^{-1} = B_b^{-1}(\mathbf{A}_b,\mathbf{\Sigma}_b)$ and $z_b^\mathrm{T} = z_b^\mathrm{T}(\mathbf{A}_b)$. The present study correlates these coefficients to the physical properties of the individual scatterers in random media. To do so effectively, we perform experiments on a model system with microscopic scatterers, whose optical properties are well defined.

In our experiments we measure $\mathbf{\Delta}_b (z)$ and its uncertainties $\sigma^2_b(z)$ by Mueller matrix ellipsometry \cite{Fujiwara_2007} and subsequent differential decomposition of the matrices \cite{Azzam_1978, Ossikovski_2011, Agarwal_2015}. We do this for uniform colloidal suspensions of spherical polystyrene (PS) particles with diameter $2r=1.5~\si{\micro\metre}$ at different volume fractions $\phi$ and a wide range of sample thicknesses $z$ illuminated in transmission by low-coherence sources at various wavelengths $\lambda$ (see Methods). Volume fractions as large as $9.6\%$ are probed which still allow neglecting optical short range interactions \cite{Fraden_1990, Saulnier_1990, Rojas_Ochoa_2004}. Our system is then fully characterized with the addition of the refractive indices of the scatterers $n_p$ and solvent $n_s$.

The spherical symmetry of the scatterers reduces the system description to only three variables $(x,~m,~\phi)$, with the size parameter $x = \frac{2\pi n_{\mathrm{s}}r}{\lambda}$ and relative refractive index $m = \frac{n_{\mathrm{p}}}{n_{\mathrm{s}}}$ \cite{Bohren_2007}. A change in the wavelength $\lambda$ is equivalent to an inverse change in the particle radius $r$. Consequently, our results can be extended to $(\lambda,r,n_\mathrm{s},n_\mathrm{p})$ that lead to similar $(x,~m)$.

Colloidal suspensions of PS spheres have several advantages \cite{Hunter_2012}. They are easy to control experimentally and their single-particle scattering is described through Mie-theory, from which the length scales $\ell$ and $\ell^*$ are computed as functions of $(x,~\phi,~m)$ \cite{Rojas_Ochoa_2002}.  Any change in these variables will lead to different evolutions of $\mathbf{p}_b(z)$. Due to their symmetry, the scatterers depolarize exclusively by multiple scattering. The closeness of the densities of water and PS spheres suppresses sedimentation over the course of the experiment. The thermodynamic behaviour of the beads is also well understood. We make use of known relationships between $\phi$ and the structure factor during the computation of $(\ell, \ell^*)$ to account for inter-particle correlations \cite{Jackson1998, Wertheim_1963}. Furthermore, polystyrene spheres of $1.5~\si{\micro\metre}$ exhibit minuscule optical absorption and a large scattering cross-section in the visible spectrum.

Large single particle scattering cross-sections are crucial in order to access both depolarization regimes (transient and stationary) experimentally. Attenuation increases as a function of $z$ and constrains the availability of intensity to a limited $z$-range. Particle sizes with the largest scattering cross-sections maximize the therein generated depolarization.

\begin{figure}[ht]
  \centering
  \includegraphics[width=\linewidth]{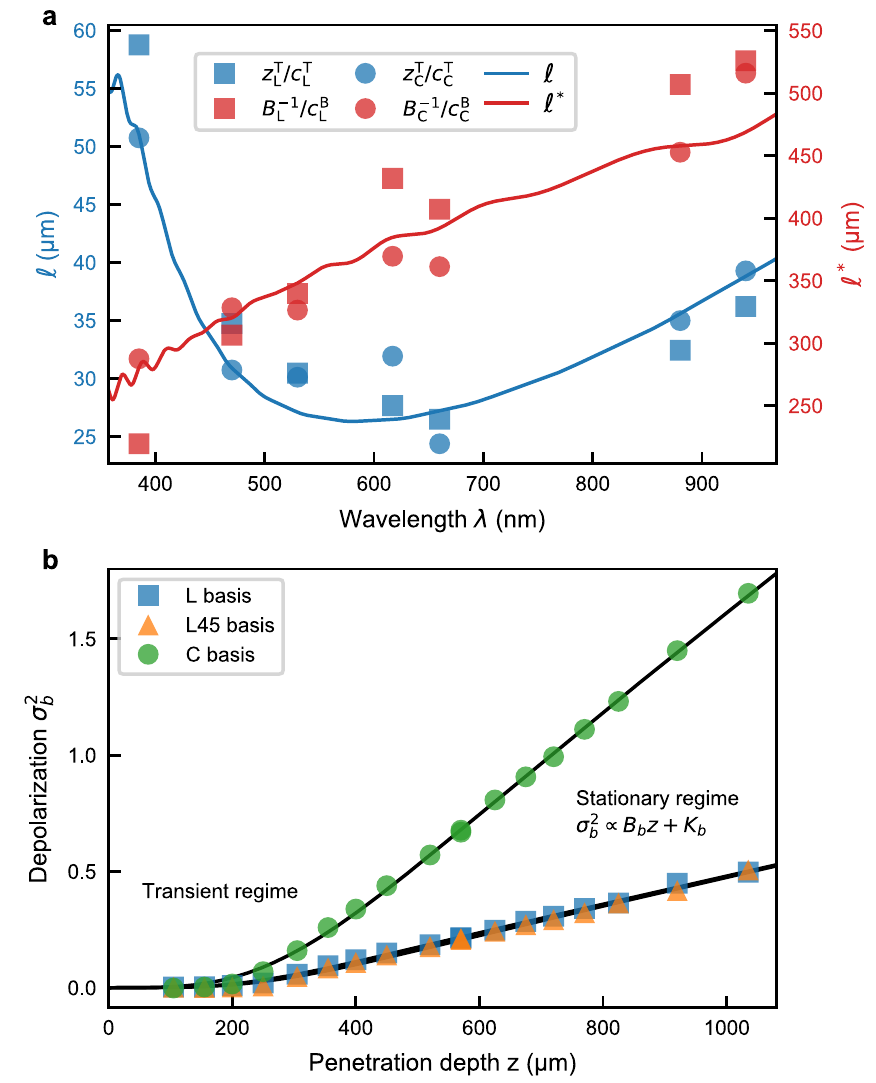}
  \caption {
    \textbf{a}, Light transport quantities (solid lines) $\ell$ (blue) and $\ell^*$ (red) show distinct spectra. Measured depolarization length scales $z_b^\mathrm{T}$ (blue symbols) and $B_b^{-1}$ (red symbols) follow $\ell$ and $\ell^*$ for linear (squares) and circular (circles) polarization (eq. (\ref{eq:zTzaB})).
 \textbf{b}, Measured depolarization (symbols) with corresponding fits to eq.~\eqref{eq:stoch} (solid lines) as a function of sample thickness $z$ for a wavelength $\lambda = 530~$nm. Each data set belongs to one of the bases (linear L, linear L45, and circular C). The transient regime is characterized by $z_b^\mathrm{T}$. The diffusion coefficient $B_b$ gives rise to the linear growth of depolarization in the stationary regime. Its inverse $B_b^{-1}$ is the corresponding depolarization length. All graphs are drawn at a volume fraction of $\phi = 1.1\%$ and sphere size $2r = 1.5~\si{\micro\metre}$.}
  \label{fig:Fig2}
\end{figure}
We want to express light depolarization in terms of transport parameters $\ell$ and $\ell^*$. It is important to differentiate their signatures in spectra of $z_b^\mathrm{T}$ and $B_b^{-1}$. This is satisfied for the chosen size of PS beads because $\ell$ and $\ell^*$ exhibit distinct spectral behaviours. While the mean free path $\ell$ displays a minimum at $600~\mathrm{nm}$, due to a Mie resonance, the transport mean free path $\ell^*$ increases monotonically with the wavelength as presented in Fig.~$\ref{fig:Fig2}\mathrm{(a)}$.

Our suspensions of $1.5~\si{\micro\metre}$ PS spheres generate noise that matches the Gaussian process described by eqns. (\ref{eq:stoch},~\ref{eq:Delta}). All measured Mueller matrices are diagonal, i.e. they are perfect depolarizers and produce maximum entropy \cite{Puentes_2005, Aiello_2005}. For this reason, an incident Stokes vector $\mathbf{S}^{\mathrm{in}} = ({\mathrm{S}_0}^\mathrm{in},{\mathrm{S}_1}^\mathrm{in},{\mathrm{S}_2}^\mathrm{in},{\mathrm{S}_3}^\mathrm{in})^\mathrm{T}$ produces the outgoing Stokes vector $\mathbf{S}^{\mathrm{out}}(z)$ that reads \cite{Charbois_2016}:
\begin{equation}
\mathbf{S}^{\mathrm{out}}(z) = I(z) \left(\begin{array}{c}
    {\mathrm{S}_0}^\mathrm{in}\\
    {\mathrm{S}_1}^\mathrm{in}\exp[- \sigma_\mathrm{L45}^2(z) - \sigma_\mathrm{C}^2(z)]\\
    {\mathrm{S}_2}^\mathrm{in}\exp[- \sigma_\mathrm{L}^2(z) - \sigma_\mathrm{C}^2(z)]\\
    {\mathrm{S}_3}^\mathrm{in}\exp[- \sigma_\mathrm{L}^2(z) - \sigma_\mathrm{L45}^2(z)]\\
  \end{array} \right),
  \label{eq:Stokes}
\end{equation}
with the transmitted irradiance $I(z)$, the Stokes parameters $\mathrm{S}_{0-3}$ and the depolarizations $\sigma_b^2$ (eq. \eqref{eq:sigmaDelta}) associated with polarization bases $i \in \{\mathrm{L},\mathrm{L45},\mathrm{C} \}$. The mean values $\langle \mathbf{\Delta_b}(z) \rangle$ vanish and do not contribute in eq. \eqref{eq:Stokes} because the Mueller matrices are diagonal. We assume that the non-zero variances $\sigma_b^2(z)$ are predominant compared to higher order statistical moments, as is confirmed in the following, where we demonstrate the accuracy of the model.

Good agreement is observed between measurements and fits of $\sigma_b^2(z)$ as shown in Fig.~$\ref{fig:Fig2}\mathrm{(b)}$ for the three optical properties L, L45 and C, at chosen $(\phi=1.1\%,~\lambda=530~\mathrm{nm})$. The stochastic model perfectly reproduces the expected transient and stationary regimes of the data. Over the transient regime, $0 \lesssim z < z^\mathrm{T}_b$, the variance $\sigma_b^2(z)$ evolves non-linearly, and subsequently enters the stationary regime, $ z^\mathrm{T}_b < z$, where it approaches the linear trend of eq. \eqref{eq:BK}. Excellent correspondence between data and model is observed for all measured $(\lambda,\phi)$.

We experimentally varied the wavelength over $\lambda \in [385~\mathrm{nm};~940~\mathrm{nm}]$ and recovered the spectra of $z_b^\mathrm{T}$ and $B_b^{-1}$ shown in Fig. \ref{fig:Fig2}(a) (see Methods). Here, clear differences between the two quantities is seen. While $z_b^\mathrm{T}$ exhibits a minimum around $\lambda=600~\mathrm{nm}$, $B_b^{-1}$ increases monotonically with $\lambda$ for all polarization bases. They follow $\ell$ and $\ell^*$, whose spectra are overlaid with associated colors. This relationship persists when the volume fraction $\phi \in [0.48\%;~9.6\%]$ is varied. It is observed that all length scales decrease monotonically with increasing $\phi$.

\begin{figure}[htb]
  \includegraphics[width=89mm]{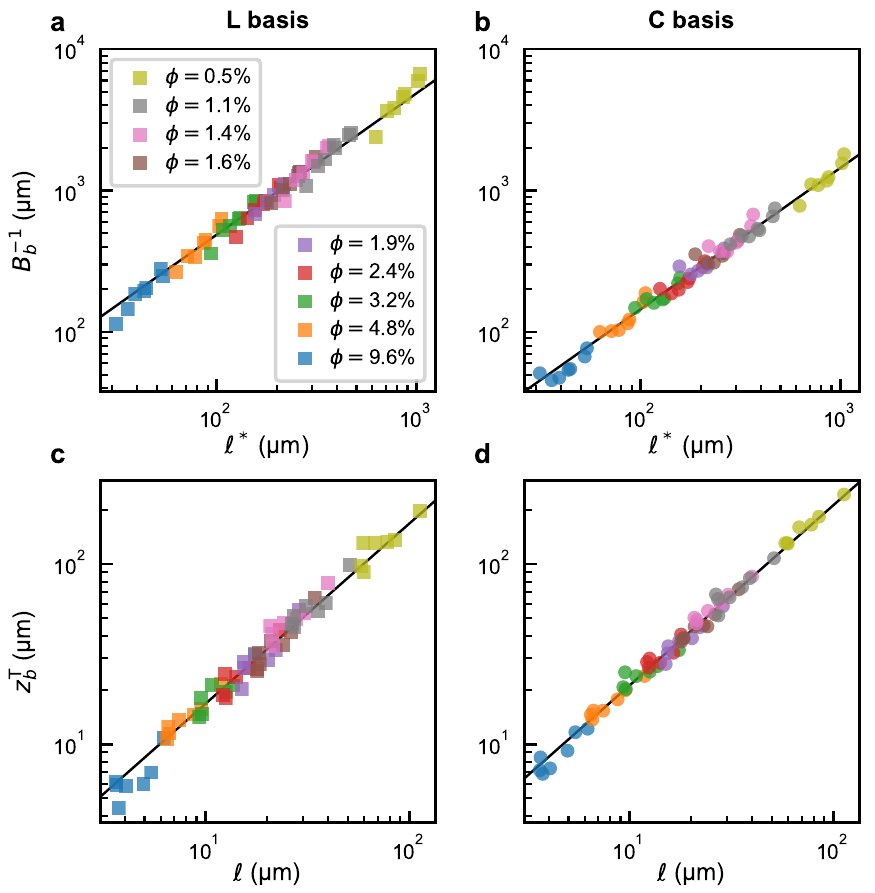}
  \caption{Depolarization length $B^{-1}_{b}$ (\textbf{a}, \textbf{c}) and transient length $z^{\mathrm{T}}_{b}$ (\textbf{b}, \textbf{d}) plotted as functions of $\ell^*$ and $\ell$, respectively. Data points were obtained from parameters $\mathbf{A}_{b}$ and $\mathbf{\Sigma}_{b}$ using fits to eq.~\eqref{eq:stoch}, while $\ell$ and $\ell^*$ were computed from Mie theory using the known values of ($\phi$, $x$). Each color corresponds to a different volume fraction of scatterers measured at different wavelengths. The left and right columns correspond to linear (L) and circular (C) polarization, respectively.}
  \label{fig:Bza}
\end{figure}

To investigate the apparent relationship between these length scales, in Fig.~\ref{fig:Bza} all measured data for $z^{\mathrm{T}}_{b}$ and $B^{-1}_{b}$ are plotted as functions of $\ell$ and $\ell^*$, respectively. All points follow straight lines evidencing a simple proportionality between the length scales.
Hence, we obtain the following relationships:
\begin{equation}
  \begin{aligned}
    z^{\mathrm{T}}_b(\ell) &= c^{\mathrm{T}}_b  \ell(\lambda,\phi), \\
    B^{-1}_b(\ell^*) &= c^{\mathrm{B}}_b  \ell^*(\lambda,\phi),
    \label{eq:zTzaB}
  \end{aligned}
\end{equation}
with fitted dimensionless coefficients $c_b^{\mathrm{T}}$ and $c_b^{\mathrm{B}}$ reported in Tab.~\ref{tab:fitcoeff}. The direct proportionality holds over nearly two orders of magnitude with no additional contributions.

This result directly links depolarization to the properties of the scattering medium. Explicit formulas are well-established for $(\ell, \ell^*)$ \cite{Rojas_Ochoa_2002}. By substitution into eqns. \eqref{eq:zTzaB} they generate explicit formulas for $(z^{\mathrm{T}}_{b}, B^{-1}_{b})$ in terms of the medium properties $(x, \phi)$. This not only enables quantitative predictions of depolarization, but also offers new avenues to measure $(\ell, \ell^*)$. Polarization measurements allow to determine $\ell$ and $\ell^*$ using eqns. (\ref{eq:Delta}, \ref{eq:stoch}, \ref{eq:Stokes}, \ref{eq:zTzaB}). This provides the advantage that acquisition times much longer than the fast diffusion of light are sufficient. The following discussion highlights the wide validity of the coefficients $c_b^{\mathrm{T}}$ and $c_b^{\mathrm{B}}$ for random media.

We expect that the values of $c_b^{\mathrm{T}}$ and $c_b^{\mathrm{B}}$ are not only invariant with respect to $\lambda$ and $\phi$ but also to $r$ for a suspension of spheres and fixed illumination conditions. In our experiments the parameters $(x, \phi, m)$ fully characterize length scales for multiple scattering \cite{Bohren_2007, Rojas_Ochoa_2002} and depolarization. The wavelength and particle radius contribute equally in the ratio $x$ and do not occur elsewhere. Consequently, a change of wavelength $\lambda$ is equivalent to an inverse change of radius $r$ and results in the same $x$ and $(\ell, \ell^*)$. Therefore, the observed invariance of the coefficients $(c_b^{\mathrm{T}},~c_b^{\mathrm{B}})$ with respect to $x$ include $\lambda$ and $r$.

\begin{table}[htb]
  \centering
  \ra{1.4}
  \begin{tabular}{cccc@{}}
    \toprule
    & L & L45 & C \\
    \cmidrule(lr){2-2} \cmidrule(lr){3-3} \cmidrule(l){4-4}
    $c^{\mathrm{T}}_{b}$ & $1.68 \pm 0.05$ & $1.70  \pm 0.06$ & $2.13  \pm 0.04$ \\
    $c^{\mathrm{B}}_{b}$ & $4.88 \pm 0.14$ & $4.85 \pm 0.13$ & $1.45 \pm 0.04$ \\
    \bottomrule
  \end{tabular}
  \caption{Linearity coefficients $(c_b^{\mathrm{T}},~c_b^{\mathrm{B}})$ between real and polarization spaces as given by eqns.~\eqref{eq:zTzaB}, fit from Fig.~\ref{fig:Bza} and given with $95 \%$ confidence intervals (linear L, linear L45, and circular C).}
  \label{tab:fitcoeff}
\end{table}
With eq. \eqref{eq:zTzaB} there is experimental evidence for a deep physical connection between the randomization of the propagation direction of light $\mathbf{k}$ and depolarization. Our results indicate that, in random media, it is impossible to restrict diffusion to only the polarization or the real space. Both domains appear inseparable and the coefficients $(c_b^{\mathrm{T}},c_b^{\mathrm{B}})$ give measures for how the rates of their dynamics compare to each other. For example, $c_{\mathrm{L}}^{\mathrm{T}}$ gives a transient regime $z_{\mathrm{L}}^{\mathrm{T}}$ lasting $\sim 1.7$ scattering events (Tab.~\ref{tab:fitcoeff}). After this distance the depolarization dynamics approach their long-term steady-state behaviour. Here, with a factor of $c_{\mathrm{L}}^{\mathrm{B}} \approx 4.9$, the rate $1/\ell^*$ at which the $\mathbf{k}$-vector is disoriented is considerably higher than the rate of depolarization $B_{\mathrm{L}}$. Diffusion of circularly polarized light through the medium is slower compared to linearly polarized light, because $c_{\mathrm{C}}^{\mathrm{B}}$ is smaller than $c_{\mathrm{L}}^{\mathrm{B}}$ and $c_{\mathrm{L45}}^{\mathrm{B}}$.

Finally, the phenomenological model of eq. \eqref{eq:stoch} reveals details about physical processes behind the depolarization dynamics. A relationship between the drift matrix $\mathbf{A}_b$ and $\ell$ is established upon substitution of $z_b^{\mathrm{T}}(\mathbf{A}_b) = 1/\sqrt{\mathrm{det}(\mathbf{A}_b)}$ into eq. \eqref{eq:zTzaB}:
\begin{equation}
  \mathrm{det}(\mathbf{A}_b) = \left(c_b^{\mathrm{T}}\ell\right)^{-2}.
  \label{eq:Aell}
\end{equation}
The drift matrix $\mathbf{A}_b$ encodes the deterministic dynamics that connect past and present values of $\mathbf{p}(z)$ (eq. \eqref{eq:stoch}). At the same time its inverse determinant quantifies the resilience of the polarization during its diffusion in the medium. Therefore, a small determinant protects the state from polarization noise over longer distances. According to eq. \eqref{eq:Aell}, the deterministic processes are characterized by $\ell$ and dominate the dynamics at the beginning of the medium. They are the reason for the non-linear evolution of $\sigma_b^2(z)$ in the transient regime $0 \leq z \leq z_b^{\mathrm{T}}$.

In summary, the two parameters $z^{\mathrm{T}}_{b}$ and $B^{-1}_{b}$ define natural length scales of depolarization. We show that, multiplied by the appropriate constants, they equal light transport variables. The resulting coefficients $c^{\mathrm{T}}_{b}$ and $c^B_{b}$ (Tab. \ref{tab:fitcoeff}) are constants for all measured wavelengths $\lambda$ and volume fractions $\phi$ and are also invariant with respect to particle sizes $r$. Their values quantify how fast the diffusion dynamics happen in the real and polarization spaces. The universality of our relations might be extended by varying the refractive indices $n_{\mathrm{p}}$ and $n_{\mathrm{s}}$ or the particle shapes \cite{Bohren_2007, Rojas_Ochoa_2002}.

\section{Methods}
\paragraph{\textbf{Colloidal samples}}
Uniform colloidal suspensions are prepared at various volume fractions $\phi$ between $0.5\%$ and $9.6\%$ by mixing pure water with a commercial master solution (microparticles.de) of $\phi = 9.6\%$. The suspension consists of polystyrene spheres of diameter $1.5\pm 0.05~\si{\micro\metre}$. Interparticle correlations and their influence on $\ell$ and $\ell^*$ are negligible at these concentrations \cite{Hunter_2012}. The particle density (1.05 g/cm$^3$) is similar to that of water and no sedimentation was observed over the duration of experiments.

\begin{figure}[h]
  \includegraphics[width=\linewidth]{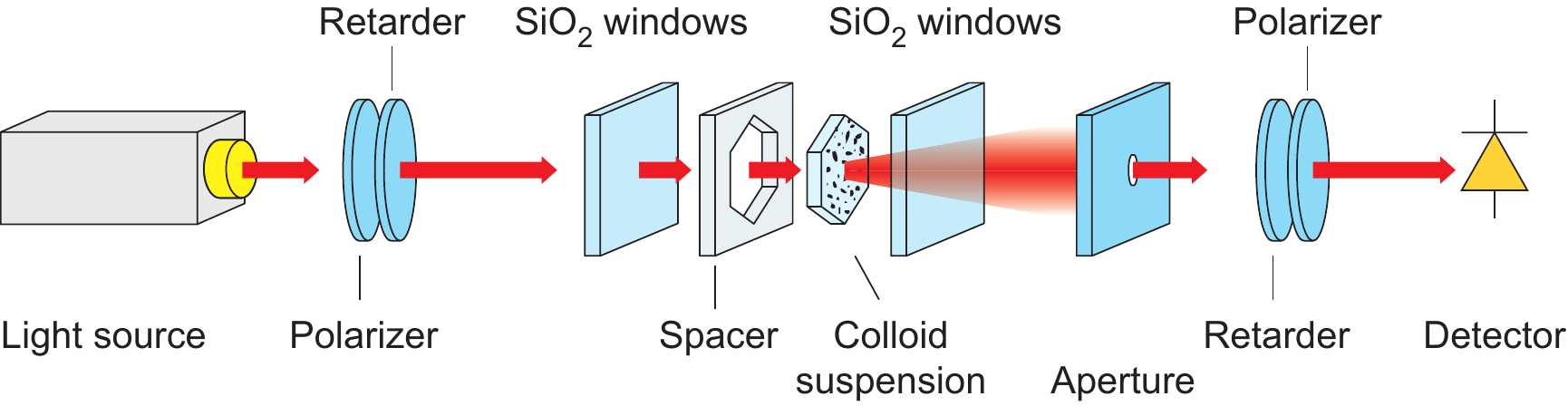}
  \caption{Schematic of the experimental setup. Depolarization is obtained from measured Mueller matrices. They are acquired from collimated LED light whose polarization is prepared from a polarizer and retarder. Light then crosses through a colloidal suspension of polystyrene beads over a thickness adjused by a PTFE spacer. The outgoing diffuse ligth is vignetted before analysis by retarder and polarizer with the resulting intensity measured by a photodiode.}
  \label{fig:setup}
\end{figure}
\paragraph{\textbf{Experimental setup}}
For measurements, the suspensions are injected via a syringe into a commercial fluid cell (Omni Cell, Specac), whose thickness $z = 12-2000~\si{\micro\metre}$ is adjusted by a variety of PTFE spacers. To fully determine their optical response the $z$-dependent Mueller matrices are acquired in the visible and near-infrared from 385 to 940~nm at normal incidence and room temperature with a Woollam VASE ellipsometer equipped with additional compensators to acquire all 16 Mueller matrix elements. To provide sufficient optical intensity, high power LEDs with low coherence are employed as light sources and with spectral bandwidths of $\Delta\lambda \sim 10~$nm to avoid speckles. The incident beam is collimated to ensure a pure polarization state $\mathbf{S}_\mathrm{in}$ and has a diameter of 5 mm while the emerging diffuse beam is detected by a 5 mm aperture. Reproducibility is successfully verified by repeatively measuring specific data points $(\lambda, \phi)$ on different days and samples. A schematic of the setup is shown in Fig.~\ref{fig:setup}.

The Brownian motion of the particles influences the experimental Mueller matrices on time scales of $t_\mathrm{acq} \sim 0.1~$s, which is much smaller than the measurement time $t_\mathrm{acq} \sim 30~$s. As a consequence, the detector captures a large ensemble of light paths. However, we stress, that instead of being detrimental to the results, this effect ensures better statistics due to the consideration of a larger ensemble of particle configurations. In the analysis, this is taken into account by the averaging operator $\langle...\rangle$ as discussed in eq. \eqref{eq:sigmaDelta}.
\newline
\paragraph{\textbf{Acknowledgments}}
G.S. thanks Thibault Chervy for fruitful discussions. The work was supported by the Deutsche Forschungsgemeinschaft DFG via DR228/38-1.
\paragraph{\textbf{Data availability}}
All data needed to evaluate the conclusions in the paper are present in the paper. 
\paragraph{\textbf{Competing interests}}
The authors declare no competing interests.
\paragraph{\textbf{Author contributions}}
G.S. conceived the study. M.G. performed the measurements. M.G. and G.S. assessed the data. G.S., M.G. and B.G. wrote the manuscript. The study was supervised by B.G. and M.D.

\onecolumngrid
\appendix

\section{Supplementary Note 1: Mean free paths}
In ensembles of spherical scatterers the mean free paths $\ell$ and $\ell^*$ depend on individual and collective properties of the particles \cite{Rojas_Ochoa_2002}:
\begin{align}
  \begin{aligned}
    \ell\ &= \left(\frac{3\phi}{2k_\mathrm{s}^2r^3}\int_0^{2k_\mathrm{s}}\frac{dC_\mathrm{sca}}{dq}F(q,\phi) q dq\right)^{-1}, \\
    \ell^* &= \left(\frac{3\phi}{4 k_\mathrm{s}^4 r^3}\int_0^{2k_\mathrm{s}}\frac{dC_\mathrm{sca}}{dq} F(q,\phi) q^3 dq\right)^{-1}.
    \label{eq:lls}
  \end{aligned}
\end{align}
Here, $k_\mathrm{s}$ is the $\mathbf{k}$-vector inside the solvent, $r$ the radius of the spheres and $\phi$ their volume fraction. The integration is performed numerically. The differential scattering cross-section $dC_\mathrm{sca}/d\Omega$ follows from Mie theory and is given by the series \cite{Bohren_2007}
\begin{align}
\frac{dC_\mathrm{sca}}{d\Omega} = \frac{2\pi}{k_\mathrm{s}^2} \left\lvert \sum_{l=0}^\infty\sqrt{2l+1}\left( a_l\mathbf{X}_{l,\pm 1} \pm i b_l \hat{\mathbf{r}} \times \mathbf{X}_{l,\pm 1} \right)\right\rvert ^2,
\label{eq:MiedCsca}
\end{align}
which converges sufficiently around the order
\begin{align}
  l_\mathrm{max} = \mathrm{floor}(k_\mathrm{s} r + 4.05 (k_\mathrm{s} r)^{1/3} + 2).
\end{align}
We use the following definition for the vector spherical harmonics \cite{Jackson1998}
\begin{align}
  \mathbf{X}_{l,m} = \frac{\widehat{\textbf{L}}Y_{l,m}(\theta,\phi)}{\sqrt{l(l+1)}},
\end{align}
in terms of the standard spherical harmonics $Y_{l,m}$ and angular momentum operator $\widehat{\textbf{L}}$. The coefficients $a_{l}$ and $b_{l}$ read:
\begin{align}
  \begin{aligned}
a_{l} &= \frac{ j_l(kr)\frac{d}{du}[u j_l(k_\mathrm{s} u)]_{u=r} - \frac{k_\mathrm{p}}{k_\mathrm{s}}\frac{\mu_\mathrm{p}}{\mu_\mathrm{s}}j_l(k_\mathrm{p}r)\frac{d}{du}[u j_l(k_\mathrm{s} u)]_{u=r} }{ \frac{k_\mathrm{p}}{k_\mathrm{s}}\frac{\mu_\mathrm{p}}{\mu_\mathrm{s}}j_l(k_\mathrm{p}r)\frac{d}{du}[u h^{(1)}_l(k_\mathrm{s} u)]_{u=r} - h^{(1)}_l(k_\mathrm{s}r)\frac{d}{du}[u j_l(k_\mathrm{p} u)]_{u=r} } \\
  b_{l} &= \frac{ \frac{\mu_\mathrm{p}}{\mu_\mathrm{s}}j_l(k_\mathrm{s}r)\frac{d}{du}[u j_l(k_\mathrm{s} u)]_{u=r} - \frac{k_\mathrm{p}}{k_\mathrm{s}}j_l(k_\mathrm{p}r)\frac{d}{du}[u j_l(k_\mathrm{s} u)]_{u=r} }{ \frac{k_\mathrm{p}}{k_\mathrm{s}}j_l(k_\mathrm{p}r)\frac{d}{du}[u h^{(1)}_l(k_\mathrm{s} u)]_{u=r} - \frac{\mu_\mathrm{p}}{\mu_\mathrm{s}}h^{(1)}_l(k_\mathrm{s}r)\frac{d}{du}[u j_l(k_\mathrm{p} u)]_{u=r}},
  \end{aligned}
\end{align}
with $j_l(x)$ the spherical Bessel functions and $h^{(1)}_l(x)$ spherical Hankel functions of the first kind. Here $\mu$ is the magnetic permeability and the subscripts ``p'' and ``s'' denote quantities inside the particle and the solvent, respectively.

The mean free paths $\ell$ and $\ell^*$ (eq. (\ref{eq:lls})) use the Percus-Yevick approximation \cite{Wertheim_1963} for the static structure factor of hard spheres given by
\begin{align}
  F(q,\phi)  = \frac{1}{1-\overline{N}C(q,\phi)},
\end{align}
where
\begin{align}
  \begin{aligned}
    \overline{N}C(q,\phi) = -24\phi &\left(\lambda_1\left[ \frac{\sin(2qr)-(2qr)\cos(2qr)}{(2qr)^3} \right] \right. \\
    &\left. - 6\phi\lambda_2 \left[ \frac{((2qr)^2-2)\cos(2qr)-2(2qr)\sin(2qr)+2}{(2qr)^4} \right] \right. \\
    &\left. -\phi \frac{\lambda_1}{2 (2qr)^6} \left[ ((2qr)^4 -12(2qr)^2 + 24)\cos(2qr) + (24(2qr)-4(2qr)^3)\sin(2qr) -24 \right] \right).
  \end{aligned}
  \label{eq:Fq}
\end{align}
This expression depends on the volume fraction $\phi$, the spheres radius $r$ and the parameters $\lambda_1$, $\lambda_2$ which are defined as
\begin{align}
  \lambda_1 = \frac{(1+2\phi)^2}{(1-\phi)^4}\quad \mathrm{and} \quad \lambda_2 = \frac{-(1+\phi/2)^2}{(1-\phi)^4}.
\end{align}
The structure factor $F(q,\phi)$ is based on the statistical mechanics of hard sphere colloids and characterizes their spatial correlations.
Within our experimental ranges for wavelengths $\lambda$ and volume fractions $\phi$, contributions from the structure factor $F(q,\phi)$ only weakly perturbs the behaviour of $\ell$ and $\ell^*$.

\section{Supplementary Note 2: Expression for the $\mathbf{\Delta(z)}$ variance}
The 2-element vector $\mathbf{\Delta}_b(z)$ describes polarization changes in the birefringence-dichroism space in one of the three possible polarization bases $b \in \{\mathrm{L},\mathrm{L45},\mathrm{C}\}$. The rate $\mathbf{p}_b(z) = \frac{d}{dz}\mathbf{\Delta}_b(z)$ satisfies the stochastic differential equation given by eq. (\ref{eq:stoch}) and has the formal solution \cite{Charbois_2016}
\begin{align}
  \mathbf{p}_b(z) = e^{\mathbf{A}_bz}\mathbf{p}_b(z_0) +  \int_{z_0}^z e^{\mathbf{A}_b(z-r)}\mathbf{\Sigma}_bd\mathbf{N}(r).
  \label{eq:psolution}
\end{align}
Our system of colloidal spheres does not polarize at $z=0$. Therefore, we set
\begin{align}
  \mathbf{\Delta}_b(z_0) = \mathbf{p}_b(z_0) = 0,~\mathrm{and}~ z_0 = 0.
\end{align}
Through the use of It$\hat{\mathrm{o}}$'s isometry \cite{Charbois_2016} we obtain the covariance matrix of $\mathbf{p}_b$
\begin{align}
\begin{aligned}
\mathrm{Cov}[\mathbf{p}_b(u),\mathbf{p}_b(v)] = \langle \mathbf{p}_b(u)\mathbf{p}_b^\mathrm{T}(v) \rangle = \int_{0}^{\mathrm{min}(u,v)} e^{\mathbf{A}_b(u-r)}\mathbf{\Sigma}_b\mathbf{\Sigma}_b^\mathrm{T}e^{\mathbf{A}_b^\mathrm{T}(v-r)}dr
\end{aligned}
\label{eq:Covp}
\end{align}
and the covariance matrix of $\mathbf{\Delta}_b$
\begin{align}
\begin{aligned}
\mathrm{Cov}[\mathbf{\Delta}_b(s),\mathbf{\Delta}_b(t)] = \int_0^sdu\int_0^tdv\int_{0}^{\mathrm{min}(u,v)} e^{\mathbf{A}_b(u-r)}\mathbf{\Sigma}_b\mathbf{\Sigma}_b^\mathrm{T}e^{\mathbf{A}_b^\mathrm{T}(v-r)}dr.
\end{aligned}
\label{eq:CovD}
\end{align}
The experimentally accessible depolarization $\sigma_b^2(z)$ is the trace of eq. (\ref{eq:CovD}) evaluated at $s=t=z$, which permits the following simplifications:
\begin{align}
\begin{aligned}
{\sigma}_b^2(z) &= \mathrm{Tr} \left\{ \int_0^z du \int_0^z dv \int_0^{\mathrm{min}(u,v)}e^{\mathbf{A}_b(u-r)}\mathbf{\Sigma}_b\mathbf{\Sigma}_b^\mathrm{T}e^{\mathbf{A}_b^\mathrm{T}(v-r)}dr \right\} \\
&=  \int_0^z du \int_0^z dv \int_0^{\mathrm{min}(u,v)} \mathrm{Tr} \left\{ e^{\mathbf{A}_b(u-r)}\mathbf{R}^\mathrm{T}\mathbf{Q}_b\mathbf{R}e^{\mathbf{A}_b^\mathrm{T}(v-r)} \right\}dr \\
&= \int_0^z du \int_0^z dv \int_0^{\mathrm{min}(u,v)} \mathrm{Tr} \left\{ \mathbf{R}^\mathrm{T}e^{\mathbf{\Gamma}_b(u-r)}\mathbf{R}\mathbf{R}^\mathrm{T}\mathbf{Q}_b\mathbf{R}\mathbf{R}^\mathrm{T}e^{{\mathbf{\Gamma}_b}^\mathrm{T}(v-r)}\mathbf{R} \right\} dr \\
&= \int_0^z du \int_0^z dv \int_0^{\mathrm{min}(u,v)} \mathrm{Tr} \left\{ e^{\mathbf{\Gamma}_b(u-r)}\mathbf{Q}_be^{{\mathbf{\Gamma}_b}^\mathrm{T}(v-r)} \right\}dr.
\end{aligned}
\label{eq:model2}
\end{align}
The product $\mathbf{\Sigma}_b\mathbf{\Sigma}_b^\mathrm{T}$ is symmetric and positive semi-definite which allows its diagonalization by a rotation matrix $\mathbf{R}(\beta)$ with angle $\beta$ and $\mathbf{Q}_b = \mathrm{diag}[q_{b,1}^2,q_{b,2}^2]$. To eliminate $\mathbf{R}$, we first transform $\mathbf{A}_b$ into a new matrix $\mathbf{\Gamma}_b = \mathbf{R}\mathbf{A}_b\mathbf{R}^\mathrm{T}$. Then we use the trace invariance with respect to cyclic permutations of factors and the identity $\mathbf{RR}^\mathrm{T} = \mathbb{1}$.
This simplification is desirable because it reduces the parameter space to six independent variables $(\mathbf{\Gamma}_{b,11},\mathbf{\Gamma}_{b,12},\mathbf{\Gamma}_{b,21},\mathbf{\Gamma}_{b,22},q_{b,1}^2,q_{b,2}^2)$. Note, that the rotation angle $\beta$ does not appear in the final expression of eq. (\ref{eq:model2}). The $2 \times 2$ matrices $\mathbf{\Gamma}_b$ and $\mathbf{Q}_b$ are determined through fits. Although the drift matrix $\mathbf{A}_b$ can not be fully recovered, its properties such as its determinant and trace are shared by $\mathbf{\Gamma}_b$. The two matrices are related through an unknown similarity transformation $\mathbf{\Gamma}_b = \mathbf{R}\mathbf{A}_b\mathbf{R}^\mathrm{T}$.

The rotational invariance of the expression for $\sigma_b^2(z)$ with respect to $\mathbf{R}(\beta)$ implies that the six recovered parameters in general describe the evolution of $\mathbf{p}_b(z)$ in coordinates that are arbitrarily rotated with respect to birefringence-dichroism coordinates. Out of the of the 8 degrees of freedom given by the parameters of $\mathbf{A}_b$ and $\mathbf{\Sigma}_b$ the experimentally observed evolution for $\sigma_b^2(z)$ has two that are not directly measurable.

Initial fits showed that the eigenvalues of $\mathbf{A}_b$ are always complex conjugates. In the following this is used to further simplify the expression of eq. (\ref{eq:model2}) with the introduction of the real $\gamma_b / 2$ and imaginary $\omega_b / 2$ parts of the eigenvalues of $\mathbf{A}_b$. The six parameters of the model then read $(\gamma_b$, $\omega_b$, $\mathbf{\Gamma}_{b,12}$, $\mathbf{\Gamma}_{b,21}$, $q_{b,1}^2$, $q_{b,2}^2)$. With
\begin{align}
\begin{aligned}
\gamma_b &= \mathrm{Tr}(\mathbf{A}_b) = \mathbf{A}_{b,11} + \mathbf{A}_{b,22} = 2\mathrm{Re}(\mathrm{eig}(\mathbf{A}_b)) \\
\omega_b &= i\sqrt{(\mathbf{A}_{b,11} - \mathbf{A}_{b,22})^2 + 4\mathbf{A}_{b,12}\mathbf{A}_{b,21}} = 2\mathrm{Im}(\mathrm{eig}(\mathbf{A}_b)),
\end{aligned}
\label{eq:eigA}
\end{align}
and after integration of eq. (\ref{eq:model2}) the model of $\sigma_b^2(z)$ has the following general form:
\begin{align}
\begin{aligned}
\sigma_b^2(z) = B_bz + K_b &+ [D_b\cos(\omega_b z/2) + E_b\sin(\omega_b z/2) ]e^{\gamma_b z/2} \\
 & + [G_b\cos(\omega_b z) + H_b\sin(\omega_b z) + I_b]e^{\gamma_b z}.
\end{aligned}
\label{eq:model3}
\end{align}
The coefficients have the following explicit expressions:
\begin{align}
\begin{aligned}
B_b = \frac{-2}{(\gamma_b^2 + \omega_b^2)^2} &[ (q_{b,1}^2 + q_{b,2}^2)(\omega_b^2-\gamma_b^2) - 4(\mathbf{\Gamma}_{b,12} - \mathbf{\Gamma}_{b,21})(\mathbf{\Gamma}_{b,12}q_{b,2}^2 - \mathbf{\Gamma}_{b,21}q_{b,1}^2 ) \\
& + 2\gamma_b(q_{b,1}^2 - q_{b,2}^2)(- \omega_b^2 - 4\mathbf{\Gamma}_{b,12}\mathbf{\Gamma}_{b,21})^\frac{1}{2} ],
\end{aligned}
\label{eq:B_b}
\end{align}

\begin{align}
\begin{aligned}
K_b = \frac{1}{\gamma_b(\gamma_b^2 + \omega_b^2)^3} &[6(q_{b,1}^2 + q_{b,2}^2)(\gamma_b^2 - 3\omega_b^2)\gamma_b^2 + 4(11\gamma_b^2 - \omega_b^2)(\mathbf{\Gamma}_{b,12}-\mathbf{\Gamma}_{b,21})(\mathbf{\Gamma}_{b,12}q_{b,2}^2 - \mathbf{\Gamma}_{b,21}q_{b,1}^2) \\
& - 6(q_{b,1}^2 - q_{b,2}^2)\gamma_b(3\gamma_b^2 - \omega_b^2)(- \omega_b^2 - 4\mathbf{\Gamma}_{b,12}\mathbf{\Gamma}_{b,21})^\frac{1}{2} ],
\end{aligned}
\end{align}

\begin{align}
\begin{aligned}
D_b = \frac{-1}{(\gamma_b^2 + \omega_b^2)^3} &[(8\gamma_b^3 - 24\gamma_b\omega_b^2)(q_{b,1}^2 + q_{b,2}^2) + (q_{b,2}^2 - q_{b,1}^2)(24\gamma_b^2 - 8\omega_b^2)(- \omega_b^2 - 4\mathbf{\Gamma}_{b,12}\mathbf{\Gamma}_{b,21})^\frac{1}{2} \\
 &+ 64\gamma_b(\mathbf{\Gamma}_{b,12} - \mathbf{\Gamma}_{b,21})(\mathbf{\Gamma}_{b,12}q_{b,2}^2 - \mathbf{\Gamma}_{b,21}q_{b,1}^2)],
\end{aligned}
\end{align}

\begin{align}
\begin{aligned}
E_b = \frac{1}{\omega_b(\gamma_b^2 + \omega_b^2)^3} &[(8\omega_b^4 - 24\gamma_b^2\omega_b^2)(q_{b,1}^2 + q_{b,2}^2) + 32(\mathbf{\Gamma}_{b,12} - \mathbf{\Gamma}_{b,21})(- \mathbf{\Gamma}_{b,21}q_{b,1}^2 + \mathbf{\Gamma}_{b,12}q_{b,2}^2)(\gamma_b^2 - \omega_b^2) \\
 &+ (24\gamma_b\omega_b^2 - 8\gamma_b^3)(q_{b,1}^2 - q_{b,2}^2)(- \omega_b^2 - 4\mathbf{\Gamma}_{b,12}\mathbf{\Gamma}_{b,21})^\frac{1}{2} ],
\end{aligned}
\end{align}

\begin{align}
\begin{aligned}
G_b = \frac{1}{\omega_b^2(\gamma_b^2 + \omega_b^2)^3} &[(q_{b,1}^2-q_{b,2}^2)(2\omega_b^4 - 6\gamma_b^2\omega_b^2)(- \omega_b^2 - 4\mathbf{\Gamma}_{b,12}\mathbf{\Gamma}_{b,21})^\frac{1}{2} + (2\gamma_b^3\omega_b^2 - 6\gamma_b\omega_b^4)(q_{b,1}^2 + q_{b,2}^2) \\
&+ (12\gamma_b\omega_b^2 - 4\gamma_b^3)(\mathbf{\Gamma}_{b,12} - \mathbf{\Gamma}_{b,21})(\mathbf{\Gamma}_{b,12}q_{b,2}^2 - \mathbf{\Gamma}_{b,21}q_{b,1}^2)],
\end{aligned}
\end{align}

\begin{align}
\begin{aligned}
H_b = \frac{-1}{\omega_b(\gamma_b^2 + \omega_b^2)^3} &[(2\omega_b^4 - 6\gamma_b^2\omega_b^2)(q_{b,1}^2+q_{b,2}^2) + (12\gamma_b^2 - 4\omega_b^2)(\mathbf{\Gamma}_{b,12} - \mathbf{\Gamma}_{b,21})(\mathbf{\Gamma}_{b,12}q_{b,2}^2 - \mathbf{\Gamma}_{b,21}q_{b,1}^2) \\
 &+ (6\gamma_b\omega_b^2 - 2\gamma_b^3)(q_{b,1}^2 - q_{b,2}^2)(- \omega_b^2 - 4\mathbf{\Gamma}_{b,12}\mathbf{\Gamma}_{b,21})^\frac{1}{2} ],
\end{aligned}
\end{align}

\begin{align}
\begin{aligned}
I_b = \frac{4}{\gamma_b\omega_b^2(\gamma_b^2 + \omega_b^2)}(\mathbf{\Gamma}_{b,12} - \mathbf{\Gamma}_{b,21})(\mathbf{\Gamma}_{b,12}q_{b,2}^2 - \mathbf{\Gamma}_{b,21}q_{b,1}^2).
\end{aligned}
\end{align}
We define the depolarization length scales as follows. From eq. (\ref{eq:model3}) it is evident that all terms except $B_bz + K_b$ decay exponentially when $\gamma < 0$ for large penetration depths $z$. This condition is satisfied throughout our data and leads to the expression eq. (\ref{eq:BK}) and the identification of $B_b^{-1}$ as the stationary depolarization length (eq. (\ref{eq:B_b})). The eigenvalues of $\mathbf{A}_b$ (eq. (\ref{eq:eigA})) define the rate of decay for the non-linear terms in eq. (\ref{eq:model3}). Hence, the length of the transient regime is proportional to $z_b^{\mathrm{T}}(\mathbf{A}_b) = 1/\sqrt{\mathrm{det}(\mathbf{A}_b)} = 2 / \sqrt{\gamma^2 + \omega^2}$.

Note that $\sigma_b^2(z)$ of eq. (\ref{eq:model3}) is positive semidefinite for $z > 0$ due to its derivation from the covariance matrix in eq. (\ref{eq:CovD}). Therefore, the depolarization curve automatically passes through the origin. With eq. (\ref{eq:BK}) this implies that $B_b$ is strictly positive. Propagation in the stationary regime can only increase, not decrease depolarization.


\section{References}
\bibliography{depollength}

\end{document}